\documentclass[review]{elsarticle}

\usepackage{hyperref}

\journal{Computer Physics Communications}

\usepackage{graphicx}
\usepackage{dcolumn}
\usepackage{bm}
\usepackage{braket}
\usepackage{xcolor}
\usepackage{amsmath}
\usepackage{natbib}

\usepackage[letterpaper, margin=1in]{geometry}

\begin{document}

\begin{frontmatter}

\title{Toward Scalable Many-Body Calculations for Nuclear Open Quantum Systems using the Gamow Shell Model}

\author{N.~Michel}
\ead{nicolas.michel@impcas.ac.cn}
\address{Michigan State University, East Lansing, MI 48824, USA}
\address{School of Physics, Peking University, Beijing 100871, China}
\address{GANIL, CEA/DSM - CNRS/IN2P3, BP 55027, F-14076 Caen Cedex, France}
\address{Institute of Modern Physics, Chinese Academy of Sciences, Lanzhou 730000, China}

\author{H.M.~Aktulga}
\address{Dept. of Computer Science, Michigan State University, East Lansing, MI 48824, USA}

\author{Y.~Jaganathen}
\address{Institute of Nuclear Physics PAN, ul. Radzikowskiego 152, Pl-31342 Krak{\'o}w, Poland}
\address{NSCL/FRIB Laboratory, Michigan State University, East Lansing, MI 48824, USA}
\address{GANIL, CEA/DSM - CNRS/IN2P3, BP 55027, F-14076 Caen Cedex, France}

\begin{abstract}
Drip-line nuclei have very different properties from those of the valley of stability, as they are weakly bound and resonant. Therefore, the models devised for stable nuclei can no longer be applied therein. Hence, a new theoretical tool, the Gamow Shell Model (GSM), has been developed to study the many-body states occurring at the limits of the nuclear chart. GSM is a configuration interaction model based on the use of the so-called Berggren basis, which contains bound, resonant and scattering states, so that inter-nucleon correlations are fully taken into account and the asymptotes of extended many-body wave functions are precisely handled. However, large complex symmetric matrices must be diagonalized in this framework, therefore the use of very powerful parallel machines is needed therein. In order to fully take advantage of their power, a 2D partitioning scheme using hybrid MPI/OpenMP parallelization has been developed in our GSM code. The specificities of the 2D partitioning scheme in the GSM framework will be described and illustrated with numerical examples. It will then be shown that the introduction of this scheme in the GSM code greatly enhances its capabilities.
\end{abstract}

\begin{keyword}
 Configuration interaction, MPI/OpenMP hybrid parallelization, 2D partitioning
\end{keyword}

\end{frontmatter}

\section{Introduction}
\label{introduction}

The Gamow Shell Model (GSM) is an extension of the traditional nuclear shell model (SM) in the complex-energy plane to describe weakly bound and resonant nuclei (see Ref.\cite{JPG} for a review and Refs.\cite{Mg,tetraneutron,FHT_Yannen} regarding recent developments). The fundamental idea in GSM is to use a one-body basis generated by a finite range potential, called the Berggren basis, which contains bound, resonant and scattering states, so that the outgoing character of the asymptotes of weakly bound and resonant nuclei can be imposed. It allows calculation of halo states, which extend far away from the nuclear zone, and resonant states, which are unbound.
  
Since GSM is a nuclear configuration interaction (CI) model, dimensions of the Hamiltonian matrix, and therefore the computational and storage costs, increase exponentially with the number of particles in a calculation. It is obviously necessary to develop a distributed memory parallel implementation of GSM, which was done in an earlier work\,\cite{JPG}. The initial version of the parallel GSM code utilized a one dimensional (1D) partitioning scheme for the Hamiltonian matrix, and simple hybrid parallelization techniques using MPI/OpenMP libraries. This initial version was convenient to implement and is well-balanced in terms of the storage and computations associated with the Hamiltonian matrix elements. It performs relatively well when the number of computational nodes utilized is small, but it requires expensive inter-node communications for large scale computations due to the 1D partitioning scheme, and it does not make effective use of the thread parallelism available on each node. Also, basis vectors of the eigensolver used in finding the nuclear states of interest were replicated redundantly on each node. These limitations significantly hamper the efficiency of the GSM code when performing large-scale calculations. 

In this paper, we describe an implementation of the GSM code which significantly improves its performance and storage requirements for large-scale calculations. As mentioned above, GSM is essentially a shell model code; therefore our implementation benefits greatly from techniques used in another shell model code called Many-body Fermion Dynamics nuclei (MFDn), which has been optimized to run efficiently on leadership class supercomputers\,\cite{SM_JPCS,SM_JPCS2,SM_JPCS3,SM_SPMM,SM_LOBPCG}. In particular, we adopt the two dimensional (2D) checkerboard partitioning scheme of MFDn, which is a powerful technique allowing to take advantage of the symmetry of the Hamiltonian matrix while reducing the MPI communication overheads\,\cite{SM_2D}. 
However, there are important differences between the GSM code and MFDn. First, GSM uses a continuous basis, i.e.~the Berggren basis, as opposed to the discrete harmonic oscillator basis used in MFDn. This has important consequences in terms of the construction of the Hamiltonian matrix in a 2D partitioned context as the matrix is less trivially sparse. Second, a fundamental feature of SM is the separation of the large proton-neutron full model space in smaller proton and neutron subspaces, whose treatment poses no numerical problem. On the contrary, the weakly bound nuclei studied within GSM often present a large asymmetry between the number of neutrons versus the number of protons. This asymmetry can lead to very large one-body spaces and demands additional memory storage optimizations in GSM, whereas they are minute compared to the total space in the regular SM. Consequently, a memory optimization absent from SM had to be devised in GSM. This optimization deals with the storage of matrix elements between Slater determinants of the same type (only protons or only neutrons) and with the storage of uncoupled two-body matrix elements. Finally, the computation of many-body resonant states in GSM has lead to additional problems absent from SM. Indeed, contrary to the bound states targeted with SM, many-body resonant states targeted with GSM are situated in the middle of scattering states, their energies therefore being interior eigenvalues. Hence, they cannot be calculated with the Lanczos method, which is well suited to calculate extremal eigenvalues, as is the case for well-bound states. As a result, we have to use an eigensolver different than the Lanczos method, used in MFDn. We use the Jacobi-Davidson (JD) method, which can directly target the interior eigenvalues and eigenvectors. We use preconditioning and angular momentum projection techniques to ensure rapid convergence of this eigensolver.

We begin by giving an overview of SM and GSM in Sect.\,\ref{SM_generalities}. Techniques used in the construction of the Hamiltonian matrix are described in Sect.\,\ref{Data_storage}. The description of the JD eigensolver, implementation of a suitable preconditioner and use of angular momentum projection to ensure rotational invariance of the basis vectors are described in Sect.\,\ref{GSM_diagonalization}. The parallelization of the Hamiltonian construction, the JD eigensolver and GSM basis orthogonalization are described in Sect.\,\ref{H_parallelization}. Examples of MPI memory storage and computation times for two nuclear systems will be given in Sect.\,\ref{GSM_MPI_computation_examples}.

\section{Background on Shell Model Approaches}
\label{SM_generalities}

\subsection{One-body states} \label{one_body_states}
\begin{figure}
  \centering
  \includegraphics[scale=0.6]{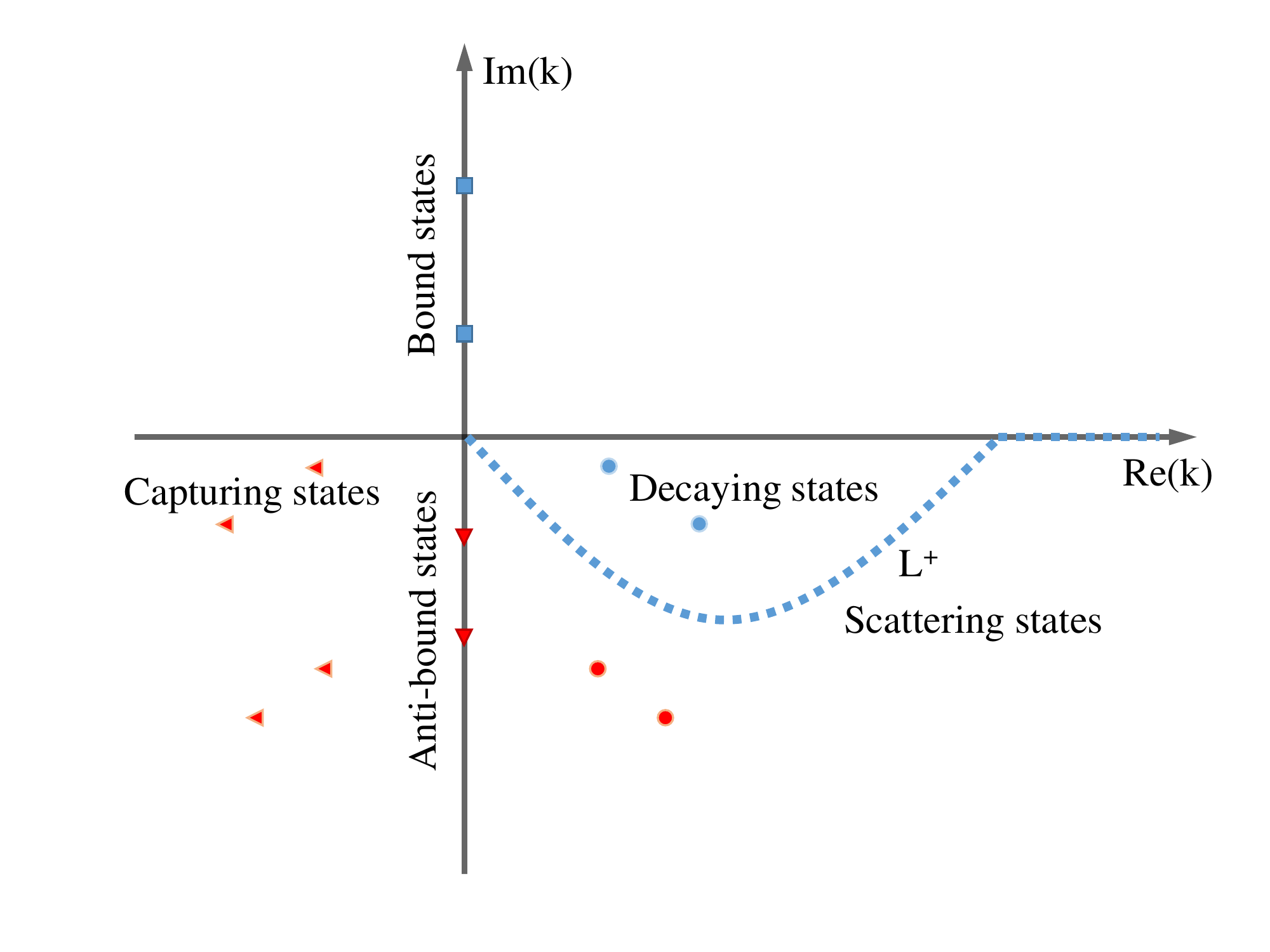}
  \caption{Berggren basis for a given partial wave. Different types of states, including bound, decaying, scattering, as well as anti-bound and capturing states, are depicted.}
  \label{Berggren_basis}
\end{figure}

The basic idea of configuration interaction (CI) is to use a basis of independent-particle many-body states to expand correlated many-body states.
In order to build  independent-particle basis states, one starts from the one-body Schr{\"o}dinger equation:
\begin{equation}
\left( \frac{\hat{p}^2}{2 \mu} + \hat{V} \right)  \ket{\phi} = e  \ket{\phi}  \label{one_body_Schrodinger_equation}
\end{equation}
where $\hat{p}^2/(2 \mu)$ is the kinetic operator, proportional to a Laplacian, $\hat{V}$ is the one-body potential, $\mu$ is the effective mass of the particle,  $\ket{\phi}$ is the one-body state solution of the one-body Schr{\"o}dinger equation and $e$ is its energy.
In most CI models, spherical potential operators $\hat{V}$ are employed \cite{SM_2D,SM_CPC} as they allow to take into account the rotational invariance of solutions exactly.
This is also well suited for the present approach where only quasi-spherical nuclei are considered.
As $\hat{V}$ is spherical, one can decompose $\ket{\phi}$ into radial and angular parts \cite{Cohen_Tannoudji}:
\begin{equation}
\phi(\vec{r}) = \frac{u(r)}{r} \mathcal{Y}^{\ell j}_m(\theta,\varphi) \label{Phi_one_body}
\end{equation}
where $u(r)$ is the radial wave function and $\mathcal{Y}^{\ell j}_m(\theta,\varphi)$ is a spherical harmonic of orbital angular momentum $\ell$ coupled to spin degrees of freedom, so that its total angular momentum is $j$ and the projection of $j$ on the $z$ axis is $m$ \cite{Cohen_Tannoudji}.
We will use the standard orbital notation in the following, where the orbital quantum number $\ell= 0$, 1, 2, 3, 4, $\dots$ is denoted as $s$, $p$, $d$, $f$, $g$, $\dots$.
Hence, for example, the angular part of a wave function bearing $\ell = 1$ and $j=3/2$ is denoted as $p_{3/2}$.
The radial wave function $u(r)$ obeys the following Schr{\"o}dinger equation:
\begin{equation}
u''(r) =  \frac{2 \mu}{\hbar^2} \left( \left( \frac{\ell(\ell + 1)}{r^2} + V_l(r) - e \right) u(r) +  \int V_{nl}(r,r')~u(r')~dr' \right) \label{Eq_ukr}
\end{equation}
where $V_l(r)$ and $V_{nl}(r,r')$ are the radial local and non-local potentials, respectively, issued from $\hat{V}$. Equation (\ref{Eq_ukr}) is solved numerically using the method of Ref.\cite{V_non_local}. 
In the traditional SM, the $\ket{\phi}$ states are harmonic oscillator states \cite{SM_2D,SM_CPC}, which are well suited for well-bound nuclear states, but not for loosely bound and resonant nuclei.
Hence, we use Berggren basis states instead \cite{Berggren}, which are generated by a finite depth potential, and contain bound, resonant and scattering states (see Fig.(\ref{Berggren_basis})). The Berggren completeness relation reads:
\begin{equation}
\sum_{n} \ket{\phi_n} \bra{\phi_n}    +  \int_{L_+}  \ket{\phi(k)} \bra{\phi(k)} \, dk = I,
\label{Berggren_comp}  
\end{equation}
where the $\ket{\phi_n}$ states are the bound states and the resonant states inside the $L^+$ contour of Fig.(\ref{Berggren_basis}). These states are usually called pole states as they are poles of the $S$-matrix \cite{JPG}. The $\ket{\phi(k)}$ states stand for scattering states and follow the  $L^+$ contour of Fig.(\ref{Berggren_basis}), and $I$ is the identity operator. 

Scattering states, which initially form a continuum, are discretized using a Gauss-Legendre quadrature with about 30 points to ensure convergence \cite{Gauss_Legendre}. Once discretized, the Berggren basis is in effect the same as that of  harmonic oscillator states in the context of CI.

As $\ket{\phi}$ has fixed  quantum numbers $j$ and $m$, angular momentum algebra can be used therein with ladder operators $j^\pm$. Let us write $\ket{\phi} =  \ket{u ~ \ell ~ j  ~ m}$ so as to make apparent the dependence on quantum numbers:
\begin{equation}
 j^\pm \ket{u ~ \ell ~ j  ~ m} = \hbar ~ \sqrt{j (j + 1) -  m (m \pm 1)}  \ket{u ~ \ell ~ j  ~ m \pm 1} \label{jpm}
\end{equation}
which has the property to raise or lower the value of $m$ by one unit keeping all other values unchanged.

Another operator based on one-body states properties is the $m$-reversal operator (also improperly called time-reversal symmetry):
\begin{equation} 
T \ket{j ~ m} = (-1)^{j - m} \ket{j ~ -m}  \label{TRS_one_body}
\end{equation}
The use of this operator allows to save memory and computations in CI, as we will see in the following sections.

\subsection{Slater determinants} \label{SD_section}
One can build independent-particle many-body states of all nucleons from antisymmetrized tensor products of $\ket{\phi}$ states of coordinates $\vec{r}$, or Slater determinants:
\begin{equation}
          SD(\vec{r}_1, \cdots, \vec{r}_A) = \sqrt{\frac{1}{A!}}
	\begin{vmatrix}
                     {\phi_1} (\vec{r}_1) & {\phi_2} (\vec{r}_1) & \cdots & {\phi_A} (\vec{r}_1) \\
                     {\phi_1} (\vec{r}_2) & {\phi_2} (\vec{r}_2) & \cdots & {\phi_A} (\vec{r}_2) \\
                     \vdots                                   & \vdots                                   & \vdots & \vdots                    \\
                     {\phi_1} (\vec{r}_A) & {\phi_2} (\vec{r}_A) & \cdots & {\phi_A} (\vec{r}_A) 
	\end{vmatrix}
\label{SD_det}
\end{equation}
where $A$ is the number of nucleons of the considered nuclear state. A complete basis can then be generated by considering all combinations of $\ket{\phi}$ states provided by Eq.(\ref{Eq_ukr}). This basis is of infinite dimension and has to be truncated in practice. For this, one considers maximal values for $\ell$ and $e$ for used $\ket{\phi}$ states, and one also limits the number of occupied scattering $\ket{\phi}$ states in Eq.(\ref{SD}) (see below), which is typically 2 to 4 at most.

Slater determinants involving pole states only are of fundamental importance, as they form the most important many-body basis states of a nuclear state decomposition. In particular, a diagonalization of the Hamiltonian in a model space generated by pole Slater determinants is called pole approximation and is used to initialize the JD method (see Sect.\,\ref{one_body_states}). Other Slater determinants are called \emph{scattering states}.

The extension of the $m$-reversal operator of Eq.(\ref{TRS_one_body}) to Slater determinants is also straightforward:
\begin{equation} 
T \ket{SD} = \prod_{i=1}^A  T_i \ket{\phi_i} \label{TRS_SD}
\end{equation}

\subsection{Occupation formalism} \label{occupation_formalism}
In the following, we will use occupation formalism \cite{Cohen_Tannoudji}. It is based on the use of creation and annihilation operators of one-body states, denoted by $a^\dagger_{\alpha}$ and  $a_{\alpha}$ respectively for the creation and annihilation of the state $\ket{\alpha}$.
Consequently, a Slater determinant can be written in a more concise form:
\begin{equation}
\ket{SD} = a^\dagger_{\phi_A}  ~ \cdots ~  a^\dagger_{\phi_1} \ket{~} = \ket{{\phi_1} ~ {\phi_2} ~ \cdots ~ {\phi_A}} \label{SD}
\end{equation}
where $\ket{~}$ is the vacuum state, which contains no particles by definition.

Occupation formalism algebra is closed and fulfill antisymmetry requirements if operators verify the following equations:
\begin{equation}
\{  a^\dagger_{\alpha} ,  a^\dagger_{\beta} \} = 0 \mbox{ , } \{  a_{\alpha} ,  a_{\beta} \} = 0 \mbox{ , } \{  a^\dagger_{\alpha} ,  a_{\beta} \} = \delta_{\alpha \beta} \label{a_dagger_a}
\end{equation}
where $\ket{\alpha}$ and $\ket{\beta}$ are one-body states and  brackets denote the anticommutation relation:
\begin{equation}
\{  O_1 ,  O_2 \} = O_1 O_2 + O_2 O_1 \label{anticommutation}
\end{equation}
where $O_1$ and $O_2$ are two operator functions of $a^\dagger$ and $a$.

It is then convenient to write the Hamiltonian in occupation formalism:
 \begin{equation}
  H = \sum_{\alpha \beta} \braket{\alpha | h |  \beta} {a^\dagger_\alpha} {a_\beta} + \sum_{\alpha < \beta , \gamma < \delta} \braket{\alpha \beta | V | \gamma \delta} {a^\dagger_\alpha} {a^\dagger_\beta} {a_\delta} {a_\gamma} \label{H_occupation_formalism}
 \end{equation}
where $h$ is the one-body part of $H$, containing its kinetic and mean-field part, while $V$ is its two-body part, embodying inter-nucleon correlations, and $\alpha$, $\beta$, $\gamma$, $\delta$ are one-body states.
In the following, as $h$ matrix elements are used with one creation operator and one annihilation operator, and as $V$ is used with two creation and two annihilation operators, they will be referred as the one particle-one hole  (1p-1h) part and  two particle-two hole (2p-2h) part, respectively.

In particular, with $N_s$ being the number of states used in the one-body basis, one can see that the number of 1p-1h matrix elements scales as $N_s^2$ and that of 2p-2h matrix elements scales as $N_s^4$. In order to provide with an order of magnitude for $N_s$, let us consider a $p_{3/2}$ contour for the Berggren basis. As the Gauss-Legendre quadrature imposed to the Berggren basis contour provides with about 30 states (see Sect.\,\ref{one_body_states}), and as one has 4 possible $m$-projections for a $p_{3/2}$ shell, one has $N_s = 120$. It is thus clear that the number of 1p-1h calculations in $H$ (see Eq.(\ref{H_occupation_formalism})) is at least four orders of magnitude smaller than the number of the 2p-2h calculations. Consequently, the 1p-1h part can be neglected from a performance point of view.

A matrix element $\braket{SD_f | H | SD_i}$, with $\ket{SD_i}$ and $\ket{SD_f}$ the initial and final Slater determinants, respectively, can be written as a function of one-body and two-body matrix elements, as well as expectation values of creation and annihilation matrix elements between $\ket{SD_i}$ and $\ket{SD_f}$ (see Eq.(\ref{H_occupation_formalism})). The latter are in particular equal to 0 or $\pm 1$ from Eq.(\ref{a_dagger_a}).

\subsection{Rotational invariance} \label{J2_projection}
The basis of Slater determinants provided by Eq.(\ref{SD}) is complete and fully antisymmetric. However, it is not rotationally invariant, as the total angular momentum is not defined therein. Its projection on the $z$ axis, however, is conserved, since its value $M$ is equal to the sum of one-body total angular momentum projections $m_i$, $i \in [1:A]$ (see Eq.(\ref{Phi_one_body})).
Consequently, given that $M$ is a good quantum number in nuclear states, we only have to consider the Slater determinants of fixed total angular momentum projection $M$ in a shell model calculation, which is called the $M$-scheme approach in CI \cite{SM_ANP}. 

This approach is preferred to $J$-scheme \cite{SM_ANP}, where Slater determinants are replaced by independent-particle many-body states coupled to $J$. Due to the conservation of the many-body total angular momentum $J$ at basis level in $J$-scheme, CI dimensions are typically smaller by a factor of 5-20 in light nuclei.
However, the $J$-scheme formulation is also more difficult to implement due to antisymmetry requirements and generally leads to denser Hamiltonian matrices (more non-zero elements)\,\cite{aktulga_jscheme}. 

The $J$ quantum number can be imposed in $M$-scheme by using appropriate linear combinations of Slater determinants. This can be effected because the $\hat{J}^2$ operator is closed in $M$-scheme, as it connects Slater determinants whose one-body states differ through their $m$ quantum number only, i.e.~belonging to the same configuration (also called partition) \cite{SM_2D,SM_CPC}.
 A configuration enumerates its occupied shells without consideration of the $m$ quantum numbers of the occupied one-body states. For example, if $\ket{0s_{1/2}(-1/2) ~ 0s_{1/2}(1/2) ~ 0p_{3/2}(-1/2)}$ is a Slater determinant, its associated configuration is $[0s_{1/2}^2 ~ 0p_{3/2}^1]$.
Hence, it is always possible to build $J$ coupled states in a configuration from linear combinations of its Slater determinants. This is done using the L{\"o}wdin operator \cite{Lowdin}:
\begin{equation}
  P_J = \prod_{J' \neq J} \frac{\hat{J}^2 - J(J+1)~I}{J'(J' + 1) - J(J+1)} \label{PJ_Lowdin}
\end{equation}
which projects out all the angular momenta $J' \neq J$. It has been checked numerically that the $J$ quantum number conservation is precise up to $10^{-10}$ or less in our applications.
In order to apply $\hat{J}^2$, one considers its standard formulation:
\begin{eqnarray}
&&\hat{J}^2 = J^- J^+ + M(M + 1)~I \label{J2} \\
&&J^\pm = \sum_{i=1}^A j^\pm_i \label{Jpm}
\end{eqnarray}
where Eq.(\ref{jpm}) has been used. As $J^\pm$ connects Slater determinants of the same configuration, and whose $M$ quantum numbers vary by one unit, it is a very sparse one-body operator, as such the computation of $\hat{J}^2$ is fast as well. 
Consequently, even though Slater determinants do not have $J$ as a good quantum number, the use of $P_J$ of Eq.(\ref{PJ_Lowdin}) allows to efficiently build $J$ coupled linear combinations of Slater determinants.

The $M$-reversal symmetry (see  Eq.(\ref{TRS_one_body}) and Eq.(\ref{TRS_SD})) is a direct consequence of rotational invariance. Indeed, rotational invariance implies that the physical properties of a shell model eigenvector are independent of $M$, while the $M$-reversal symmetry implies that they are invariant through the symmetry $M \leftrightarrow -M$.
In particular, if $M=0$, the $M$-reversal symmetry allows to calculate only half of the output shell model vector when one multiplies $H$ by an input shell model vector. Indeed, the components of the Slater determinants $\ket{SD}$ and $T \ket{SD}$ differ by a phase equal to $\pm 1$ straightforward to calculate.
Consequently, as it is always possible to calculate shell model eigenvectors of even nuclei with $M=0$, even nuclei are twice faster to calculate than originally expected.

Theoretically, the use of $P_J$ of Eq.(\ref{PJ_Lowdin}) to impose $J$ as a good quantum number is not necessary. Indeed, as $[H,\vec{J}] = 0$, a linear combination of Slater determinants coupled to $J$ provides another vector coupled to $J$ when acted on by $H$. However, due to numerical inaccuracy, $H$ and $\vec{J}$ do not exactly commute, so that the $J$ quantum number can be lost after several matrix-vector multiplications. The obvious treatment for this is to suppress the shell model vector components with $J' \neq J$, when it is no longer an eigenstate of $\hat{J}^2$, which is effected using  Eq.(\ref{PJ_Lowdin}).
In order to know whether $P_J$ has to be applied or not, one has to test whether a shell model vector is an eigenstate of $\hat{J}^2$ or not, which is rather fast. Indeed, if $M = J$, as is usually the case, shell model vectors are checked if they are eigenstates of $\hat{J}^2$ if the action of $J^+$ on this shell model vector provides zero, as the angular momentum projection of a vector coupled to $J$ cannot be larger than $J$. The only exception to this rule is when one uses $M$-reversal symmetry when $J > 0$, as $M = 0$ in all cases when $M$-reversal symmetry is applied. In this case, one has to calculate the action of $\hat{J}^2 - J(J+1)~I$ on the considered shell model vector to check if it is equal to zero. 
Hence, as one applies $J^\pm$ only once or twice after the matrix-vector operation borne by $H$ application, this test is very fast compared to $H$ or $P_J$ application.

\section{Data Storage in GSM}
\label{Data_storage}

In the SM and GSM approaches, memory utilization is of critical importance as the sizes of the matrices involved grow rapidly with the problem size. To draw a comparison between SM and GSM in this regard, let us consider $N_v$ valence nucleons in a one-body space of $N_s$ states. Based on the discussions in the previous section, one can neglect antisymmetry, parity and $M$ projection to make such a comparison, as the overall impact of these factors is minimal. Consequently, we will consider only 2p-2h matrix elements  (see Sect.\,\ref{occupation_formalism}), of the form $\braket{SD_{f} | H | SD_{i}} = \pm \braket{\alpha_{f} ~ \beta_{f} | V | \alpha_{i} ~ \beta_{i}}$. One then obtains that the dimension of the Hamiltonian in the GSM approach, which we will denote as $d$, is proportional to ${N_s}^{N_v}$ and that the probability to have a non-zero matrix element is proportional to $(1/N_s)^{N_v - 2}$, as all states must be equal in $\ket{SD_{i}}$ and $\ket{SD_{f}}$, except for $\ket{\alpha_{i} ~ \beta_{i}} \neq \ket{\alpha_{f} ~ \beta_{f}}$. This implies that the total number of non-zeros is close to $d^{1 + 2/N_v}$, which corresponds to $d^{1.67}$ and $d^{1.5}$ for 3p-3h and 4p-4h truncations in the continuum, respectively. For instance, for $d \sim 10^9$, if one compares to the typical $d^{1.4}$ number of non-zeros in SM for this dimension \cite{SM_JPCS}, the Hamiltonian for the GSM approach results in matrices which are typically one to two orders of magnitude larger than that of the regular SM. To ensure fast construction of the GSM Hamiltonian and tackle the data storage issue of the resulting sparse matrices, we have developed the techniques presented below.

\subsection{Bit algebra in SM vs state storage in GSM}

We saw that configurations and Slater determinants are necessary to build the many-body basis states of GSM. There are much fewer configurations than Slater determinants, which leads to many advantages. Due to valence space truncations, configurations are generated sequentially. This is computationally inexpensive, as there is a small number of configurations. Since Slater determinants of different configurations are independent, it is straightforward to use parallelization at this level: Configurations are distributed over available processing cores, so that all Slater determinants are generated independently by varying $m$ quantum numbers in occupied shells in a fixed configuration. The obtained Slater determinants are then distributed to all compute nodes as the full basis of Slater determinants must be present in each node. 

When building the Hamiltonian matrix $H$, we first consider the configuration and their Slater determinants. This way, searches for Slater determinant indices are very fast, as binary search is effected at the level of configurations first and at the level of Slater determinants afterwards. This also allows us to save memory space, as all indices related to shells can be stored in arrays involving configurations only, while those involving the $m$ quantum numbers only are effected in arrays involving Slater determinants.

In GSM, configurations and Slater determinants involve few valence particles, many shells and many one-body states, whereas one has many valence particles, few shells and few one-body states in SM. Thus, we use an implementation based on bit storage which is different than traditional SM \cite{SM_CPC}.
As a state can be occupied by at most one nucleon due to antisymmetry, and previously mentioned SM features, it is convenient in SM to associate a Slater determinant to a binary number. For example, $\ket{1011000000}$ is a Slater determinant where the states 1, 3 and 4 are occupied, while the states 2 and 5, ..., 10 are unoccupied. The strength of this method is that Slater determinant algebra is taken care of by operations on bits, which are very fast. However, this method becomes inefficient if one has many unoccupied states, as one integer possesses 4 bytes, or 32 bits, which can be equal to 0 or 1, so that it can contain at most 32 states. It is customary in GSM to have contours for the $s_{1/2}$, $p_{1/2}$ and $p_{3/2}$ partial waves, which are discretized with 30 points each typically (see Sect.\,\ref{one_body_states}). Conversely, it is sufficient to use 5 to 10 harmonic oscillator states for the $d_{3/2}$ and $d_{5/2}$ partial waves to obtain convergence. Let us consider as an example that we discretize contours with 30 points and that we take 10 harmonic oscillator states for the $d$ partial waves, which are in fact typical values in practice. This generates 340 one-body states, therefore 11 integers, of 44 bytes as a whole, would be necessary to store a single Slater determinant. 
This would be inefficient memory-wise because one would have to store many zeros, and maybe also inefficient performance-wise due to the larger arrays to consider.
Hence, we store configurations and Slater determinants as regular arrays, in which case the former example would be stored as $\{1,3,4\}$, requiring only 6 bytes, as each state index is represented by a short integer. 
The bit scheme and regular scheme become equivalent if one has 22 valence nucleons, which is well beyond the current capacities of our code, where the maximal number of particles in the continuum is about 4.
Calculations are fast with this storage scheme, and they can be easily loop-parallelized (unlike bit operators acting on individual bits).

\subsection{Calculation and storage of phases in GSM}

We will deal here with numerical problems generated by the large asymmetry between proton and neutron spaces.
As the problem is the same whether the neutron space is much larger than the proton space or not, 
and as we will study examples where one has valence neutrons only, we will only consider the case where the neutron space is dominant.
Obviously, proton-rich nuclei, for which the proton space is largest, bear similar properties.
As already mentioned, one cannot store all data related to the neutron space, as the space required to do so is much larger than that needed in SM.
The fundamental reason for this is the use of the Berggren basis. Indeed, one typically has 100-200 neutron valence shells, arising from the discretization of scattering contours (see Fig.(\ref{Berggren_basis})), as one has to discretize the contour of one partial wave with about 30 states, and one has 5 partial waves when $\ell \leq 2$ ($s_{1/2}$, $p_{1/2}$, $p_{3/2}$, $d_{3/2}$, $d_{5/2}$). Conversely, this number is 30 for an 8$\hbar \omega$ space in SM, as  one considers harmonic oscillator shells of the form $\ket{n ~ \ell ~ j}$ which then have to satisfy $2n + \ell \leq 8$.
Additionally, the neutron-to-proton ratio in typical GSM applications is usually very large (for instance, in the study of nuclei along the neutron drip-line) and one may often have only valence neutrons in a calculation. It would be too costly to recalculate neutron matrix elements each time. Thus, memory optimization schemes had to be devised to store matrix elements between neutron Slater determinants. They deal with what we call phases, i.e.~matrix elements involving creation or annihilation operators, only of the form $\braket{SD_f | a_\alpha^\dagger | SD_i}$, $\braket{SD_f | a_\alpha^\dagger ~ a_\beta | SD_i}$, ... that are equal to $\pm 1$, with which the indices of involved one-body shells and states, configurations and Slater determinants must be included. 

Let us define the initial and final spaces as being the spaces to which the initial $\ket{\Psi_i}$ and final $\ket{\Psi_f}$ many-body states belong, where one has $H \ket{\Psi_i} = \ket{\Psi_f}$.
In the 2D parallelization scheme (which will be discussed in more detail in Sect.\,\ref{H_parallelization}), one can use the fact that both initial and final spaces,
assigned to each processor core on the square Hamiltonian are only a fraction of the full space. They refer to the number of rows and columns, respectively \cite{SM_2D}.

Hence, we can store matrix elements of the form $\braket{SD_{int} | a_\alpha | SD_i}$ and $\braket{SD_f | a_\alpha^\dagger | SD_{int}}$, $\braket{SD_{int} | a_\alpha ~ a_\beta | SD_i}$ and $\braket{SD_f | a_\alpha^\dagger ~ a_\beta^\dagger | SD_{int}}$, 
where $\ket{SD_{int}}$ is an intermediate Slater determinant, chosen so that the data to store are minimal.
It is clear that any one-body or two-body observable can be calculated with this scheme.
Indeed, one has: 
\begin{eqnarray}
    \braket{SD_f | a_\alpha^\dagger ~ a_\beta | SD_i} &=& \braket{SD_f | a_\alpha^\dagger | SD_{int}} \braket{SD_{int} | a_\beta | SD_i} \label{one_body_phase_2D} \\
    \braket{SD_f | a_\alpha^\dagger ~ a_\beta^\dagger ~ a_\delta ~ a_\gamma | SD_i} &=& \braket{SD_f | a_\alpha^\dagger ~ a_\beta^\dagger | SD_{int}} \braket{SD_{int} | a_\delta ~ a_\gamma | SD_i} \label{two_body_phase_2D}
  \end{eqnarray}
The phase matrix, containing these matrix elements, is stored in a sparse form: 
One fixes $\ket{SD_i}$, and all $\ket{SD_f}$ varying by one or two states are generated. 
Thus, the obtained phase, but also the indices of $\ket{SD_f}$ and of associated one-body states ($\alpha$, $\beta$, $\gamma$ and $\delta$) must be stored.

In order to save memory, one loops firstly over configurations (see Sect.\,\ref{J2_projection}), so that only the configuration index of $\ket{SD_f}$ and the shell indices (functions of $n$, $\ell$, $j$, 
but not $m$) associated with the one-body states ($\alpha$, $\beta$, $\gamma$ and $\delta$) are stored, and one loops over the Slater determinants of that fixed configuration afterwards, where $m$-dependent values only are stored.
From that information, one can recover the full phase matrix.

One can see that the number of phases involving $\ket{SD_{int}}$ is proportional to $2 N_v$ for one-body phases and $N_v (N_v - 1)$ for two-body phases, with $N_v$ being the number of valence nucleons. Indeed, one has two arrays of one-body phases and two arrays of two-body phases. The storage of phase matrices is further optimized by requiring that one has $\alpha < \beta$ and $\gamma < \delta$ and leveraging the $M$-reversal symmetry. Indeed, if one applies the $M$-reversal operator of Eq.(\ref{TRS_SD}) to the Slater determinants present in $\braket{SD_{int} | a_\alpha | SD_i}$ and $\braket{SD_{int} | a_\alpha ~ a_\beta | SD_i}$, the obtained phase and associated indices can be easily recovered from the phase matrix element of the initial Slater determinants. This yields an additional memory gain of about a factor of 2 for the storage of phases at the cost of negligible additional numerical operations.

We note that memory optimization of phases described above is not utilized for the angular momentum operator $\hat{J}^2$.  Since $\hat{J}^2$ is function of $J^{\pm}$ (see Sect.\,\ref{J2_projection}), which are  very sparse operators, the number of phases used therein is much reduced compared to the Hamiltonian matrix. Decomposition of Eq.(\ref{one_body_phase_2D}) involving an intermediate Slater determinant is therefore not necessary.

\subsection{Construction and Storage of the Hamiltonian} \label{H_MEs}

The 1p-1h part of $H$ is negligible for performance purposes and hence is not considered in this section (see Sect.\,\ref{occupation_formalism}). Therefore, we concentrate firstly on the neutron 2p-2h part of $H$. From Eq.(\ref{H_occupation_formalism}), one has:
\begin{equation}
\braket{SD_f | H | SD_i} = \braket{SD_f | a_\alpha^\dagger ~ a_\beta^\dagger ~ a_\delta ~ a_\gamma | SD_i} \braket{\alpha \beta | V | \gamma \delta} \label{H_ME_2p_2h_neutron}.
\end{equation}
One then has to generate all the Slater determinants $\ket{SD_f}$ for a fixed Slater determinant $\ket{SD_i}$.
For this, one loops over all intermediate configurations $C_{int}$ and Slater determinants $\ket{SD_{int}}$ (see above). One then obtains the $\braket{SD_{int} | a_\delta ~ a_\gamma | SD_i}$ phase and associated one-body states. Using the same procedure on $\ket{SD_{int}}$, one generates $\braket{SD_f | a_\alpha^\dagger ~ a_\beta^\dagger | SD_{int}}$ phase and associated one-body states, so that the two-body phase $\braket{SD_f | a_\alpha^\dagger ~ a_\beta^\dagger ~ a_\delta ~ a_\gamma | SD_i}$ is obtained with its  one-body states. 
The two-body matrix element $\braket{\alpha \beta | V | \gamma \delta}$ comes forward from the knowledge of one-body states, so that the Hamiltonian matrix element $\braket{SD_f | H | SD_i}$ is obtained.

The treatment of the proton 2p-2h part of $H$ is mutatis mutandis the same as its neutron 2p-2h part, and its proton-neutron 2p-2h part of $H$ is very similar:
\begin{eqnarray}
\braket{SD_f | H | SD_i} &=& \braket{\alpha_p ~ \beta_n | V | \gamma_p ~ \delta_n}  \nonumber \\
    &\times& \braket{{SD_p}_{(f)} | a_{\alpha_p}^\dagger | {SD_p}_{(int)}}   \braket{{SD_p}_{(int)} | a_{\gamma_p} | {SD_p}_{(i)}} \nonumber \\ 
    &\times& \braket{{SD_n}_{(f)} | a_{\beta_n}^\dagger | {SD_n}_{(int)}} \braket{{SD_n}_{(int)} | a_{\delta_n} | {SD_n}_{(i)}}
\end{eqnarray}
where the proton and neutron character of states have been explicitly written. The computational method is otherwise the same as in the neutron 2p-2h part of $H$.


\subsection{On-the-fly calculation and partial storage of the Hamiltonian matrix} \label{on_the_fly_partial}

The discussion so far has focused on the \emph{full storage scheme} in the GSM code, where all elements of the Hamiltonian matrix are stored in memory. As the Hamiltonian matrix is always sparse, one only stores non-zero matrix elements and their associated indices. While this is ideal from a computational cost point of view as no recalculation of matrix elements is needed, memory requirements in GSM grow rapidly with the number of nucleons.
This is the main factor affecting problem size, as work vectors, even though stored on fast memory, take much less memory than Hamiltonian as only a few tens, or at worse hundreds, of them are necessary (see Sect.\,\ref{H_parallelization}).
Despite, the memory optimizations described above, the \emph{full storage scheme} is essentially \emph{total memory bound}; in other words, calculations possible with this scheme are limited by the aggregate memory space available on the compute nodes. Therefore, in addition to the full storage scheme, we have developed on-the-fly and partial matrix storage schemes for GSM. 

In the \emph{on-the-fly scheme}, the sparse matrix-vector multiplications (SpMVs) needed during the eigensolver are performed on-the-fly as the Hamiltonian is being recalculated from scratch at each eigensolver iteration. While memory requirements for the Hamiltonian matrix are virtually non-existent in this case, computation time is maximal due to repeated constructions of the Hamiltonian. 

In the \emph{partial storage} scheme, we do not store the final Hamiltonian matrix elements which is made up of two-body interaction coefficient multiplied by a phase, which form basically the whole Hamiltonian matrix up to a negligible part (see discussion in Sect.\,\ref{H_parallelization}). Indeed, the number of  two-body interaction coefficients is much smaller than the number of Hamiltonian matrix elements (see Sect.\,\ref{occupation_formalism}). Therefore, it is more efficient from the storage point of view to store the index and phase of two-body matrix elements in the Hamiltonian phase, so that the final two-body matrix element is determined by a lookup into the two-body interaction coefficient array and multiplied by its corresponding phase. Here, the most time consuming part is the search in the two-body interaction coefficient array, as the latter is very large and the two-body matrix elements considered in subsequent searches are not necessarily found to be contiguous in the interaction array. The obtained memory gain with the partial storage method compared to the full storage method is about 2.5, as two-body matrix elements are complex numbers and integers are stored instead, on the one hand,  while it is still necessary to store the indices of Slater determinants $\ket{SD_f}$ (see Sect.\,\ref{H_MEs}), on the other hand. Consequently, the partial storage scheme lies in between the full storage and the on-the-fly computation schemes, as its memory requirements (while still being significant) are not as large as the full storage scheme, but the partial storage method is faster than the on-the-fly scheme as it does not require reconstruction of the Hamiltonian from scratch.

\section{Diagonalization of the GSM matrix}
\label{GSM_diagonalization}

In SM, as one is only interested in the low-lying spectrum of a real symmetric Hamiltonian, the Lanczos method is the tool of choice. It is a Krylov subspace method which makes use of matrix times vector operations only and the low-lying extremal eigenvalues are the ones that converge first. While the same method could be used for searching the bound states in GSM, it leads to poor or no convergence at all for resonant states. This is due to the presence of numerous scattering states lying below a given resonant state in GSM, which would have to converge before the desired resonant state can converge in the Lanczos method. Consequently, we turn to a diagonalization method that can directly target the interior eigenvalues, i.e.~the JD method. Indeed, JD only involves sparse matrix times vector operations like the Lanczos method, but it also includes an approximated shift-and-invert method which allows it to directly target interior eigenvalue and eigenvector pairs.

\subsection{Complex symmetric character of the GSM Hamiltonian}
 
As GSM Hamiltonian matrices are complex symmetric, 
a standard problem mentioned in this case is the numerical breakdown of 
JD method \cite{complex_symmetric_householder}. Indeed, the Berggren metric is not the Hermitian metric, but the analytic continuation of the real symmetric scalar product. Consequently, the Berggren norm of a GSM vector can be in principle equal to zero. Then, this vector cannot be normalized and the 
JD iterations fail. However, for this to occur, one would need matrix elements whose imaginary parts are of the same order of magnitude as their real parts on one hand, and rather large off-diagonal matrix elements on the other hand. As neither of these cases are satisfied in practice, breakdown does not occur. Moreover, as the number of basis vectors is typically a few tens or hundreds, this additional memory storage and diagonalization time is negligible in our case. Consequently, the use of complex symmetric matrices does not lead to any convergence issues in GSM.

Let us note that the Lanczos method can be directly applied to the complex symmetric case if one replaces the Hermitian metric by the Berggren metric \cite{Berggren}.
But the Lanczos method performs poorly compared to the JD method because it requires hundreds of iterations if one calculates a loosely bound nuclear eigenstate, and does not even converge for resonant many-body eigenstates unless a full diagonalization is utilized. Nevertheless, the JD method requires an initial set of eigenpairs to start its iterations, and the Lanczos method can determine the GSM eigenpairs at the pole approximation level. But we note that one can generally perform a full diagonalization of the Hamiltonian matrix at the pole approximation level instead.

\subsection{Preconditioning} \label{Hprec}

Even though the use of an approximated shift-and-invert scheme ensures convergence of the JD method to the desired interior eigenvalues, this convergence can be very slow. Hence, a crucial need in GSM is to find a good preconditioner $H_{prec}$, which transforms the eigenvalue problem into a form that is more favorable for a numerical solver, essentially accelerating convergence. The simple diagonal approximation for $H_{prec}$ is not sufficient therein, as off-diagonal matrix elements are large in the pole space. However, $H$ is diagonally dominant in the scattering space. Consequently, to build an effective preconditioner, one can take $H_{prec}$ equal to $H$ itself in the pole space and equal to the diagonal of $H$ in the scattering space (the diagonal matrix elements $\braket{SD | H | SD}$ involving the Slater determinants of a fixed configuration are in fact replaced by their average therein in order to ensure that $[H_{prec},\vec{J}] = 0$). Using $H_{prec}$ as a preconditioner in JD requires computing its inverse matrix. Since the dimension of the pole space is very small (a few hundreds at most), and the rest of $H_{prec}$ is only a diagonal matrix, computing the inverse of $H_{prec}$ is inexpensive. Consequently, the chosen preconditioner has a minimal computational overhead, and by providing a reasonable approximation to $H$, it facilitates quick convergence. As the coupling to the continuum is small, with typically 70-80\% of the GSM eigenpairs residing in the pole space, the preconditioned JD method converges in typically 30 iterations per eigenpair, \emph{e.g.}, one needs 30 JD iterations to calculate the ground state only, and 60 iterations to calculate the ground state plus the first excited state.

\subsection{Implementation}
In light of the above discussion, the preconditioned JD method as implemented in the GSM code is summarized below (see Ref.\cite{Jacobi_Davidson} for a more detailed description of the JD method):
\begin{itemize}

\item Start from an approximation to the desired the GSM eigenpair, $E_i$ and $\ket{\Psi_i}$. The first eigenpair $E_0$ and $\ket{\Psi_0}$ is obtained from the pole approximation (see Sect.\,\ref{SD_section}), where the Lanczos method and full orthogonalization (due to the small size of the Hamiltonian) can be used. After the first eigenpair, each set of converged pairs provide an approximation for the next pair.

\item Calculate the residual $\ket{R_i} = H \ket{\Psi_i} - E_i \ket{\Psi_i}$.

\item Update the approximate eigenvector by solving the linear system $(H_{prec} - E_i I) \ket{\Phi_{i+1}} = \ket{R_i}$, where $H_{prec}$ is the preconditioner for $H$ (see Sect.\,\ref{Hprec}).

\item Orthonormalize $\ket{\Phi_{i+1}}$  with respect to all previous JD vectors $\ket{\Phi_j}$, $0 \leq j \leq i$, and project it onto $J$ if necessary (see Sect.\,\ref{J2_projection}).

\item Project $H$ onto the extended basis set $\ket{\Phi_j}$, $0 \leq j \leq i+1$, so that one obtains a very small complex symmetric matrix which is diagonalized exactly using standard methods. This provides with a new approximation to the eigenpair $E_{i+1}$ and $\ket{\Psi_{i+1}}$.
$\ket{\Psi_{i+1}}$ is identified from the spectrum of the diagonalized small complex symmetric matrix using the overlap method \cite{JPG}. For this, one looks for the eigenstate whose overlap with pole approximation (see Sect.\,\ref{SM_generalities}) is maximal. 
As pole Slater determinant components are always more important than scattering Slater determinant components (see Sect.\,\ref{SD_section}), this guarantees that the eigenstate obtained with the overlap method corresponds to the sought physical nuclear state.

\item Repeat the above steps until convergence, as indicated by the norm of the residual $\ket{R_i}$. 

\end{itemize}

We note that this procedure must be performed separately for each Hamiltonian eigenpair. However, one typically needs to calculate fewer than 5 eigenpairs per nucleus, the most common situations being calculation of the ground state only, or the ground state and the first excited state. Hence, it is not prohibitive to use a new JD procedure per eigenpair.

\section{Parallelization of the GSM code}
\label{H_parallelization}

The first version of the GSM code followed a 1D partitioning scheme, where the Hamiltonian $H$ was partitioned along its rows. The $H$ matrix, despite being symmetric, was stored as a full (but still sparse) matrix, and each block of rows was assigned to a different process. MPI, OpenMP versions, and the hybrid scheme combining the two of them, were implemented. This scheme is indeed convenient to implement and is well balanced for the storage of $H$ matrix elements, and is reasonably efficient when the number of processes is small, \emph{e.g.} less than 100 as is the case in a small-scale execution. Its main drawback is that to effect the sparse Hamitonian times vector operation during JD iterations, the entire input vector must be replicated on each node; this clearly generates expensive MPI communications. Consequently, the 1D partitioning approach is not scalable to thousands of cores which would be needed for accurate calculations of heavy nuclei. 

\begin{figure}
  \centering
  \includegraphics[scale=0.6]{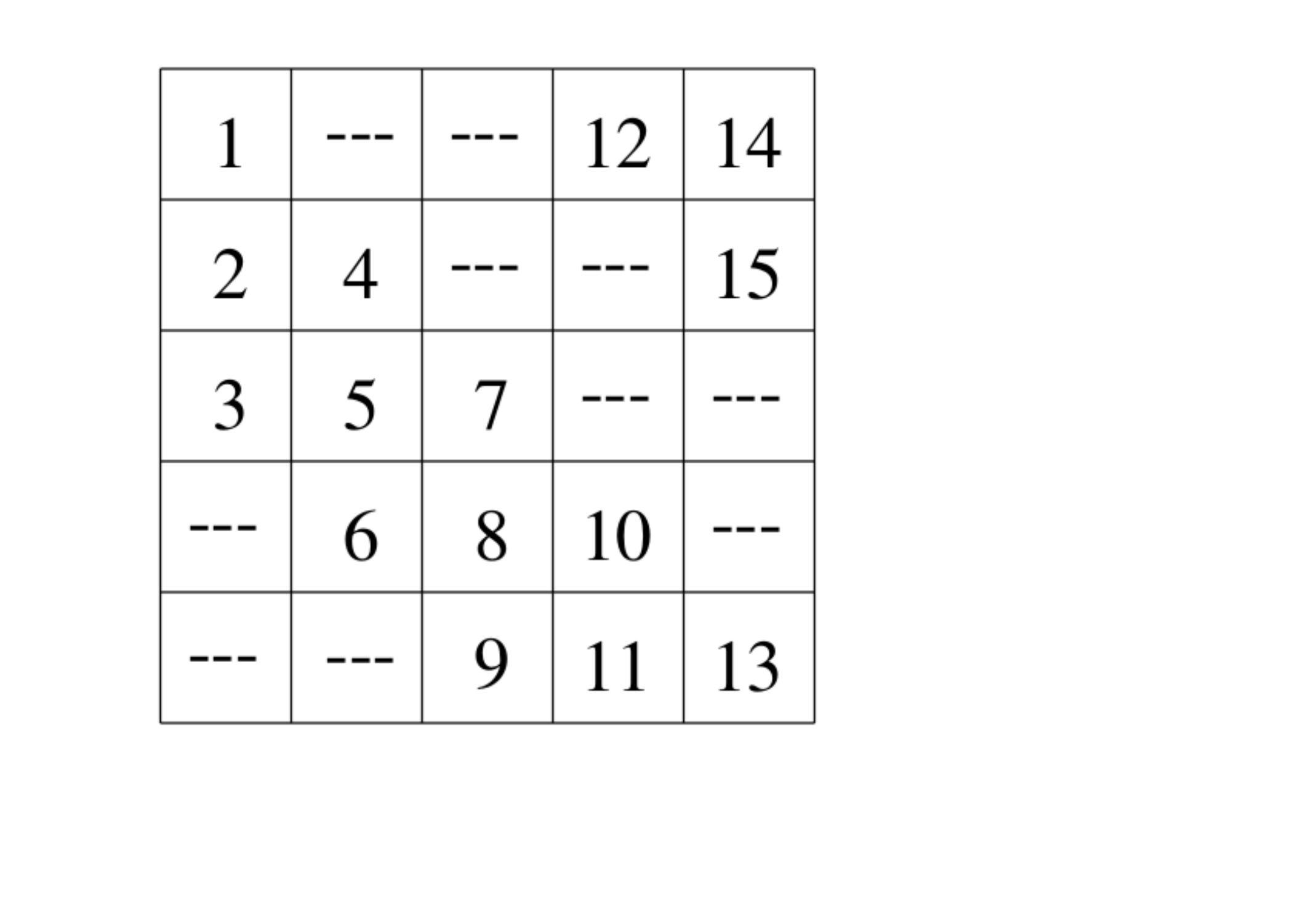}
  \vspace{-.8in}
  \caption{Example of the 2D partitioning scheme for the Hamiltonian matrix. Occupied squares are denoted by numbers and unoccupied squares by dashed lines.}
  \label{matrix_2D}
\end{figure}

In the current version of GSM, we implemented a 2D partitioning scheme that improves both memory storage and MPI communication costs of the 1D partitioning scheme.
In this scheme, the symmetry of $H$ is exploited and $H$ is divided into $N = n_d(n_d+1)/2$ squares, each of which are assigned to a different process. A simple example of the matrix distribution in the 2D scheme where $n_d=5$ and $N=15$ is depicted in Fig.(\ref{matrix_2D}). Consequently, each process receives an (almost) equal number of $\ket{SD_i}$ Slater determinants (corresponding to the \emph{rows} of the small squares assigned to them) and $\ket{SD_f}$ Slater determinants (corresponding to the \emph{columns} of the small squares assigned to them) that scale roughly with $1 / \sqrt{N}$. As the phases needed in one node are already stored therein during the construction of the Hamiltonian, each process works independently from others and there are no MPI communications at this stage. 

Compared to the 1D scheme, each process accesses a much smaller portion of the input and output vectors during the Hamiltonian times vector operations, and as a result MPI communications create a significantly smaller overhead with increasing problem dimensions. Moreover, as will be discussed in more detail below, in 2D partitioning with a hybrid MPI/OpenMP parallelization scheme, a single (or a small group of) thread(s) takes care of MPI data transfers, while all other threads are dedicated to matrix-vector multiplications. As such, MPI communications can be overlapped with matrix-vector multiplication calculations, providing even greater scalability. 

\subsection{Eigensolver Iterations}

Two main operations in eigensover iterations are i) sparse matrix times vector operations,
and ii) orthonormalization of the new JD vector with respect to the previous JD vectors.
For the $H$ times vector operation, note that as a result of exploiting symmetry of the Hamiltonian, each process must perform a regular sparse matrix multiplication (SpMV) with the submatrix it owns, and a second SpMV with the transpose of its submatrix. To perform these operations, one uses row and column communicators, which group processes along each row and column. Hence, each process is part of two groups corresponding to its own row and column, with each group having $n_d/2$ processes in them. 

\subsubsection{Distributed SpMV}
Let us now describe the algorithm of matrix times vector when one applies the Hamiltonian $H$ on a GSM vector within the 2D partitioning scheme, of dimension $d$:
\begin{itemize}
  \item All submatrices (i.e.~occupied squares in  Fig.(\ref{matrix_2D})) are effected in parallel by their owner processes.

  \item  For submatrices on the diagonal, their associated process (i.e.~the process where this submatrix is stored) is the master node of the row and column communication groups. The master node is responsible for distributing the corresponding part of the input GSM vector to all nodes on the same row and column via the row and column communicators. For example, in  Fig.(\ref{matrix_2D}), nodes 2, 4 and 15 form a row communication group, while nodes 4, 5 and 6 form a column communication group. For both communication groups, node 4 is the master node. 
  
  \item  Before any computation is performed, the master nodes broadcast their part of the input GSM vector to their column communication groups. 

  \item  Upon receiving the input vector through the column communicator, each process starts effecting the SpMV for the $H$ submatrix that it owns by utilizing multiple threads. In parallel to the SpMV, one thread takes part in the collective communication of the GSM input vectors, this time through the row communication groups in preparation for the transpose SpMV operation to follow. This way, the  SpMV operation and MPI communications are overlapped. 

\item The next step is to effect the transpose SpMV operations using the input vectors communicated through the row groups in the preceding step. While the transpose SpMV operations are being effected using multiple threads, this time the communication thread takes part in reduction of the partial outputs of the SpMV operation above to the master nodes through the row communicators, again overlapping SpMV communications with computations.

 \item Note that while performing the transpose SpMV, race conditions would occur because input and output indices are exchanged when one leverages $H$ symmetry. A block of $n_t$ vectors is used therein for output, where $n_t$ is the number of threads per node, to avoid a race conditions (see Ref.\cite{SM_2D}). These blocks of output vectors are first aggregated locally, and then are reduced at the master nodes through column communicators, thereby completing the distributed memory SpMV operation.

\end{itemize}

The cost of MPI communications is then that of $4~d/n_d$ complex numbers to be transferred collectively, with two transfers out of four being overlapped with multiplications. Consequently, this scheme is as efficient as in SM.
Moreover, given that the matrices in GSM are relatively denser than in SM (see Sect.\,\ref{Data_storage}), inter-node communications are expected to incur less overhead in GSM than in SM.

\subsubsection{GSM vectors storage and orthogonalisation} 
The JD method requires the storage of tens or hundreds of GSM vectors, which amount to about 160 Mb each already for a dimension of $10^7$ (GSM vectors are complex).To reduce the memory overhead and computational load imbalance, JD/GSM vectors are distributed among all nodes using a method similar to the 1D hierarchical partitioning scheme, first devised in  Ref.\cite{SM_2D}. In this scheme, at the end of the distributed SpMV, each master node separates their portion of the GSM vector output into (almost) equal parts and scatters them to a few processes. Consequently, the additional operations needed for vectors, in particular the reorthogonalization with respect to all previously stored JD vectors, does not pose any problem in our implementation.

\subsection{Parallelization of $\hat{J}^2$ in GSM}

We saw in Sect.\,\ref{J2_projection} that rotational invariance has to be checked through the action of $\hat{J}^2$ or $J^{+}$, and imposed if necessary with the $P_J$ operator of Eq.(\ref{PJ_Lowdin})  (see Eqs.(\ref{J2},\ref{Jpm})).

However, 2D partitioning would be rather inefficient therein due to the block structure of the $J^\pm$ matrix. Indeed, $J^\pm$ cannot connect two different configurations, which implies that it is in practice entirely contained in a diagonal square of its matrix, except for side effects. Consequently, 2D partitioning would imply that the diagonal nodes of the $J^\pm$ matrix would contain virtually all the $J^\pm$ matrix, while all the other nodes would be spectators. Thus, we choose to parallelize $J^\pm$ with a hybrid 1D/2D method, where each column of the $J^\pm$ matrix is stored on each node. GSM vectors are divided into $N$ parts, similarly to the 2D scheme, as such the output GSM vector must be reduced on each node at the end of the calculation. As each node contains a part of the diagonal squares of the $J^\pm$ matrix, the $J^\pm$ matrix memory distribution is well balanced. Symmetry requirements are here absent as $J^\pm$ is not symmetric, as it connects Slater determinants of total angular momentum projection $M$ and $M \pm 1$. As a matter of fact, the main problem here is MPI data transfer. Indeed, the number of $J'$ angular momenta to suppress (see Eq.(\ref{PJ_Lowdin})) is of the order of $50$, and, naively, one would have two MPI transfers of a full GSM vector per $J^\pm$ application, leading to $200$ MPI transfers of a full GSM vector per $P_J$ application, which is prohibitive. A solution to this problem lies in the block structure of the $J^\pm$ matrix. Indeed, the MPI transfers to be done therein are only those involving neighboring nodes in GSM vectors, so that their total volume is that of the non-zero components of a configuration in between two nodes, and not $d$, which is tractable. Moreover, $P_J$ applications occur only a few times at most, as it is rare for the $J$ quantum number not to be numerically conserved after a $H$ application, so that it does not slow down the implementation of eigenvectors significantly.

\section{Numerical Evaluations}
\label{GSM_MPI_computation_examples}

In order to test the performance of the new version of the GSM code, we consider the $^8$He and $^8$Be nuclei. Their model space consists of the $^4$He core with valence nucleons, four valence neutrons for $^8$He and two valence protons and two valence neutrons for $^8$Be nuclei. The core is mimicked by a Woods-Saxon potential and the interaction used is that of Furuchi-Horiuchi-Tamagaki type, which has been recently used for the description of light nuclei in GSM \cite{FHT_Yannen}. The model space comprises partial waves bearing $\ell \leq 3$, where the $s_{1/2}$, $p_{3/2}$, $p_{1/2}$, and $d_{5/2}$ states are given by the Berggren basis with a discretization of 21 points for contours, whereas the $d_{3/2}$, $f_{7/2}$ and $f_{5/2}$ states are of harmonic oscillator type, as in SM, where 6 states are taken into account for each of these partial waves. The model space is truncated so that no more than two nucleons can occupy scattering states and harmonic oscillator states. Proton and neutron spaces are treated symmetrically for $^8$Be. This model is used only for computational purposes, so that energies have not been fitted to their experimental values. The model spaces used nevertheless follow the usual requirements demanded for a physical calculation.

The $^8$He and $^8$Be nuclei are complementary for our numerical study. Indeed, nuclei with a large asymmetry between the number of neutrons versus the number of protons are more difficult to treat than those possessing more or less the same number of valence protons and neutrons. This is the typical case for SM, so that the on-the-fly method, based on the recalculation of matrix elements of the Hamiltonian, is usually very effective in this case. Consequently, this study also tests the ability of the code to calculate efficiently Hamiltonian matrix elements involving only neutrons, which often occurs in drip-line nuclei.

The GSM dimensions for the $^8$He and $^8$Be nuclei are respectively $d_{He}=939,033$ and $d_{Be}=3,371,395$, so that calculations remain fast with a relatively small number of nodes while remaining significant. Note that the GSM matrix are much denser than in SM, which is due to the drastic truncation used. The number of matrix non-zero elements is about $d_{He}^{1.65}$ and $d_{Be}^{1.59}$, which is about 7 and 4 times larger, respectively, than the typical number of non-zero matrix elements of SM roughly equal to $d^{3/2}$. Additionally, multiplications of complex numbers are in practice twice slower than multiplications of real numbers. Hence, even though the aforementioned dimensions would lead to very fast calculations in SM, they take a sufficiently long time in our study.

The effected parallel calculations are all hybrid as we use the MPI/OpenMP parallelization scheme. The number of nodes is of the form $n_d(n_d+1)/2$ (see Sect.\,\ref{H_parallelization}), and takes into account its possible values from 15 to 45 MPI ranks for $^8$He, and its possible values from 21 to 45 MPI ranks for $^8$Be.
Indeed, It was impossible to store to the Hamiltonian matrix of $^8$Be with the full storage method if one takes 15 nodes due to the memory limitations of a single node. 8 OpenMP threads per MPI rank are used.
All shown calculations were performed at the National Energy Research Scientific Computing Center (NERSC) of Lawrence Berkeley National Laboratory in Berkeley, CA.
Code debugging, optimization and testing were partly done using computer clusters available at  the Oak Ridge Leadership Computing Facility, as well as MSU's High Performance Computing Center (HPCC).

\begin{figure}[htbp]
\begin{tabular}{cc} 
\includegraphics[width=8cm]{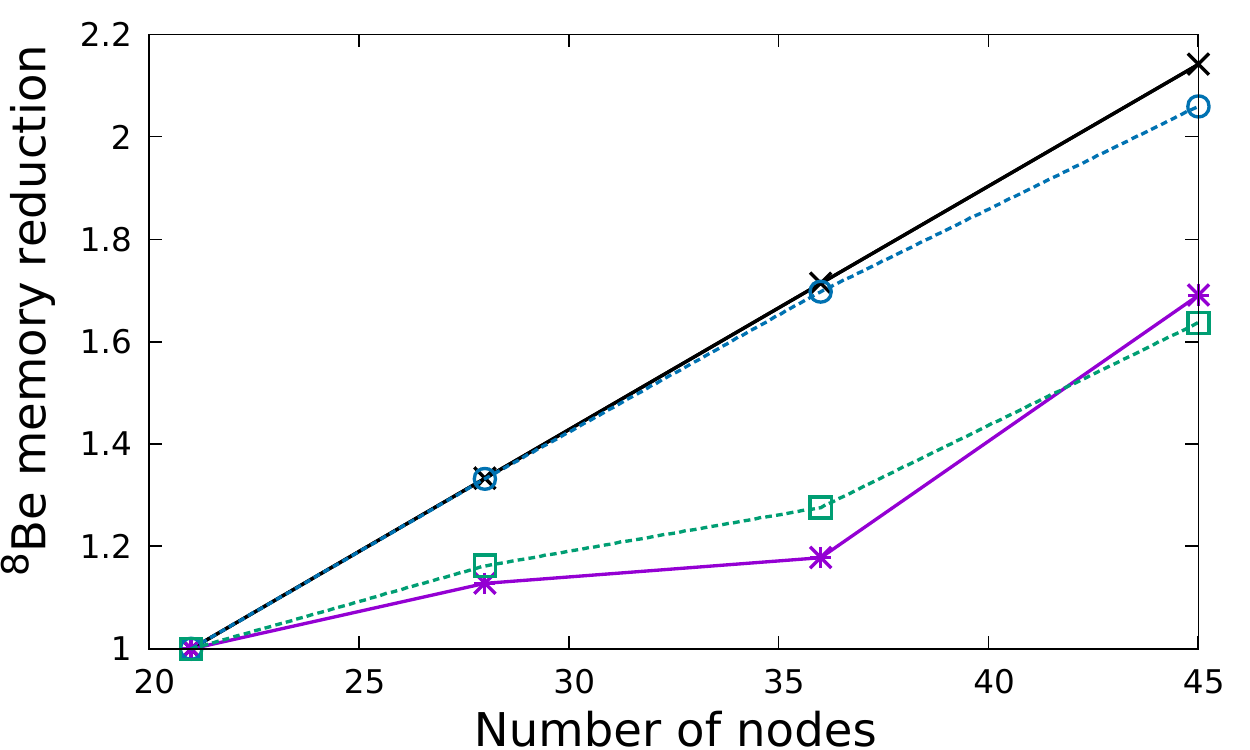} & \includegraphics[width=8cm]{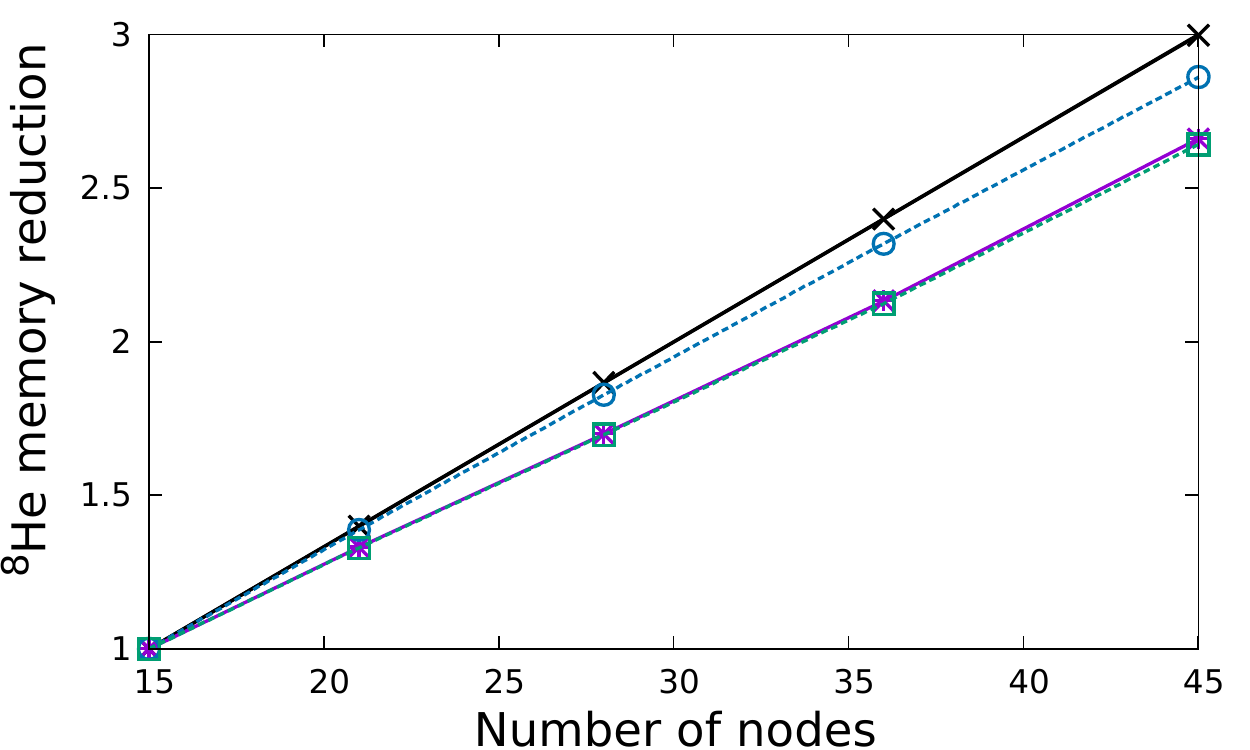}
\end{tabular}
\caption{Reduction of storage memory of Hamiltonian matrix elements per node for $^8$Be (left panel) and $^8$He (right panel) as a function of the number of nodes (color online). All data have been divided by the value obtained with the smallest number of nodes (21 for $^8$Be and 15 for $^8$He), taken as a reference point. Full storage results are represented with solid lines, with stars for maximal memory stored in a node and crosses for average memory stored in a node. Partial storage results are represented with dashed lines, with squares for maximal memory stored in a node and circles for average memory stored in a node. The strong scaling line is also depicted on the picture as a solid line, but it cannot be discerned from the depiction of average memory stored in a node obtained with the full storage method.}
\label{memory_results}
\end{figure}
Results for memory storage reduction are shown in Fig.(\ref{memory_results}),
where we show the maximum memory space required for the storage of the Hamiltonian by any one node, as well as the average space required across all nodes.
Firstly, it is clear that the average memory distribution scales very well with the number of nodes for both $^8$He and $^8$Be.
As no data of large size are stored besides the Hamiltonian in the full storage method, it is not suprising that an exact scaling is obtained if one averages over all nodes used. However, results issued from the partial storage method slightly depart from perfect scaling. This comes from the necessary additional storage of two-body matrix elements, which provide with Hamiltonian matrix elements from stored array indices and Hamiltonian phases (see Sect.\ref{on_the_fly_partial}). However, by considering the maximal memory stored in a node, one can see that scaling performance is very different for $^8$He and $^8$Be. While the maximal memory scales very satisfactorily for $^8$He, where about 90\% of perfect scaling is attained, that of $^8$Be grows unevenly and its scaling efficiency is about 75\%.
The slow increase from 28 to 36 nodes indicates a sizable load imbalance for memory distribution of the Hamiltonian matrix elements among the utilized nodes.
One can assume that it comes from a less homogeneous distribution of non-zero matrix elements in $^8$Be, generated by the presence of both valence protons and neutrons, as $^8$He only bears valence neutrons. This issue is possibly caused by a few large many-body groups ending up at the same row/column group in our relatively simple distribution scheme. It may be fixed by sorting the many-body groups based on their size before doing a round-robin distribution, or by breaking large groups into smaller ones

\begin{figure}[htbp]
\begin{tabular}{cc} 
\includegraphics[width=8cm]{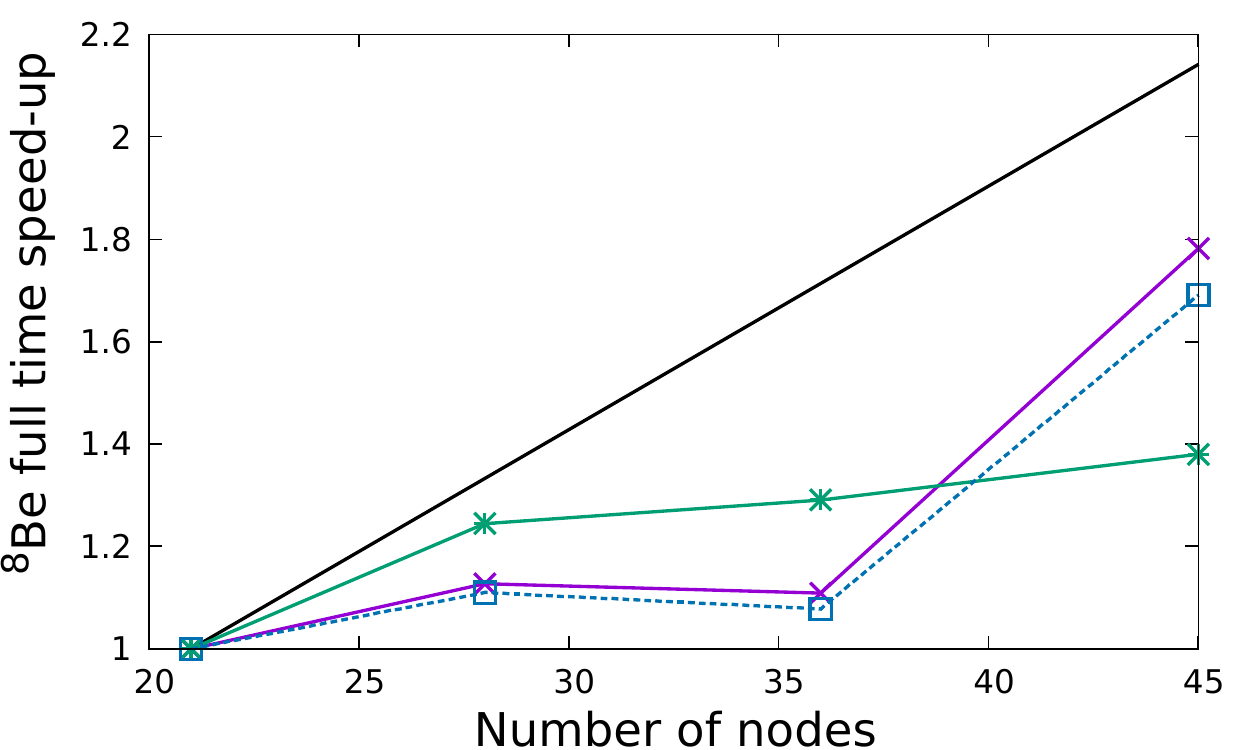} & \includegraphics[width=8cm]{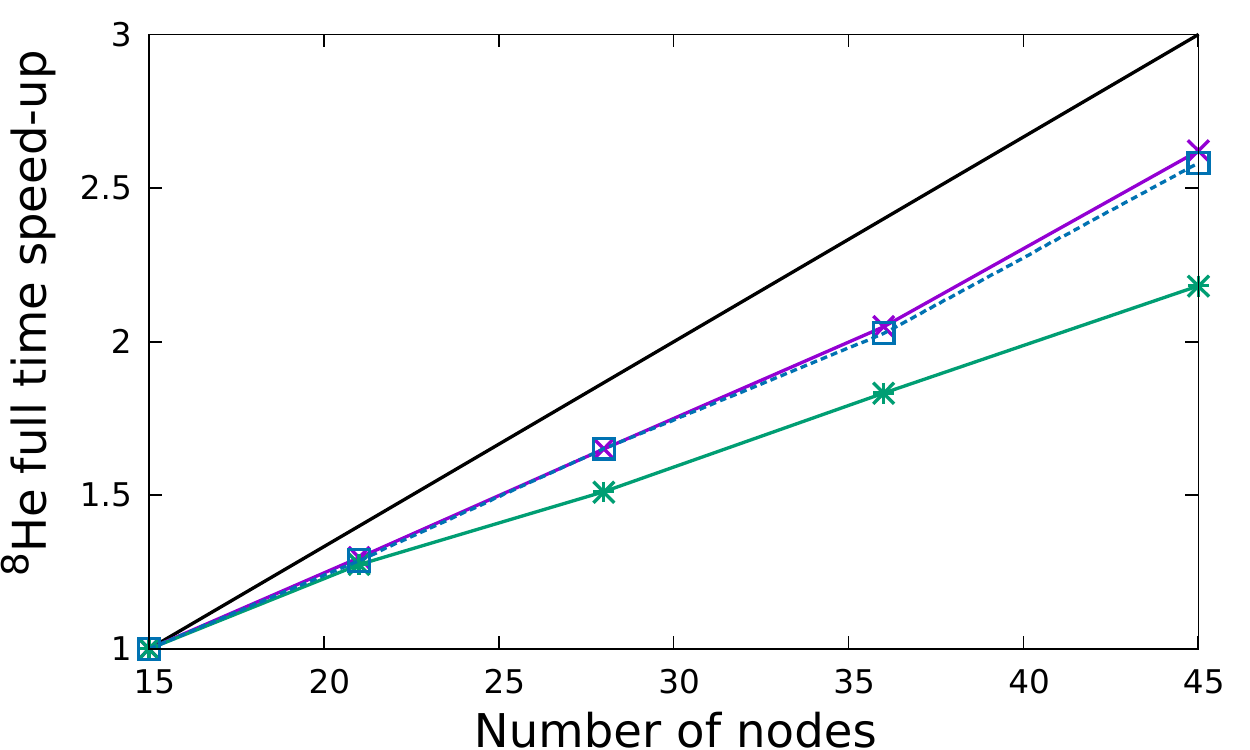}
\end{tabular}
\caption{Speed-up of the total time taken by a matrix-vector operation for $^8$Be (left panel) and $^8$He (right panel) as a function of the number of nodes (color online). All data have been divided by the value obtained with the smallest number of nodes (21 for $^8$Be and 15 for $^8$He), taken as a reference point. Full storage results are represented by a solid line with crosses, partial storage results by dashed lines with squares, and on-the-fly results by a solid line with stars. The strong scaling line is depicted on the picture as a solid line.}
\label{total_time_results}
\end{figure}
Results for the speed-up of total calculation time are shown in Fig.(\ref{total_time_results}).
Similarly to Fig.(\ref{memory_results}),
total calculation time scales very well for $^8$He, when storage and on-the-fly methods show a typical ratio of 90\% and 75\% with respect to perfect scaling, respectively, while it is not the case for that of $^8$Be.
While these performances seem to be low at first sight for $^8$Be, they follow the uneven pattern of memory distribution among nodes, provided by the full storage calculation (see the left panel of Fig.(\ref{memory_results})). Indeed, efficiency varies from 60\% to 80\% for storage methods when the number of nodes goes from 36 to 45 nodes. However, the speed-up of the on-the-fly method stagnates, as it reaches only 65\% with 45 nodes, which is significantly smaller than the values obtained with the storage methods.
Nevertheless, absolute calculation times are not excessive with the on-the-fly method compared to the partial and full storage methods,
as they are slower by a factor of about 5 and 10 for $^8$Be, respectively. 
Consequently, the on-the-fly method is of high interest when the partial and full storage methods cannot be used due to the impossibility to store the Hamiltonian matrix.
Nevertheless, its low scaling properties clearly demand to be investigated in the future.

\begin{figure}[htbp]
\begin{tabular}{cc} 
\includegraphics[width=8cm]{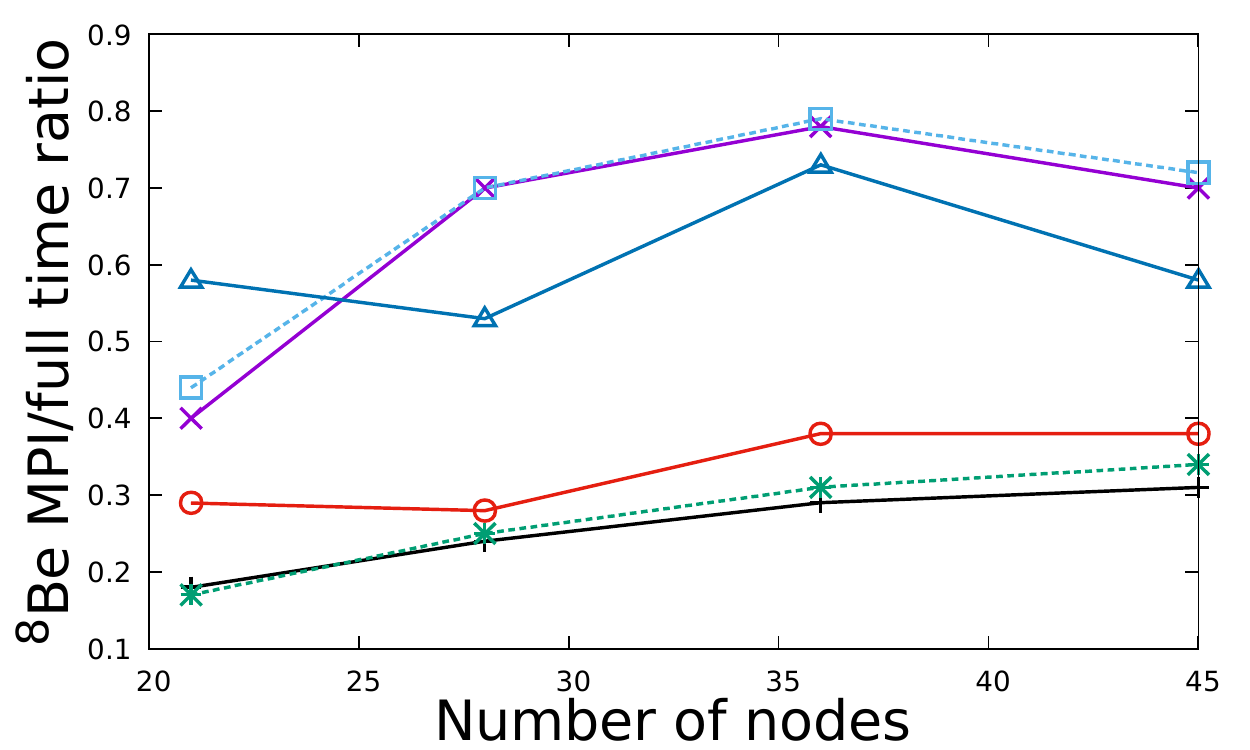} & \includegraphics[width=8cm]{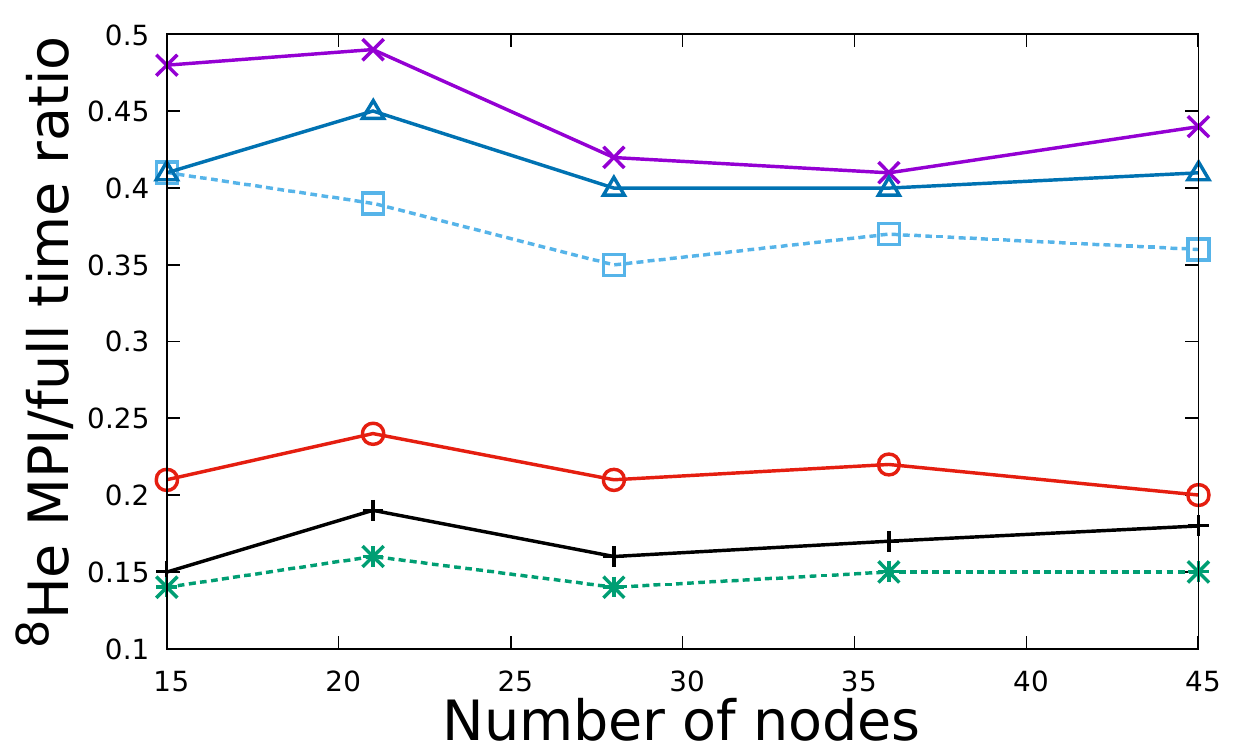}
\end{tabular}
\caption{Ratio of MPI communication time to the full time spent during a matrix-vector operation for $^8$Be (left panel) and $^8$He (right panel) as a function of the number of nodes (color online).
Full storage results are represented by solid lines, with pluses for average MPI time and crosses for maximal MPI time in a node.
Partial  storage results are represented by dashed lines, with stars for average MPI time and squares for maximal MPI time in a node.
On-the-fly results are represented by solid lines, with circles for average MPI time and triangles for maximal MPI time in a node.}
\label{MPI_time_results}
\end{figure}
Results for MPI communication times are shown in Fig.(\ref{MPI_time_results}) as the ratio of MPI communication time to the full time spent during a matrix-vector operation.
As the GSM vectors transferred with MPI are the same in full, partial and on-the-fly methods,
MPI communication times should be identical if scaling with the number the nodes were perfect. However, the situation is very different in practice.
In fact, one obtains a fairly large value for the ratio of average and maximal MPI communication times to total times, of 20\%-50\% for $^8$He and 30\%-70\% for $^8$Be (see Fig.(\ref{MPI_time_results})).
Moreover, obtained values for average and maximal MPI times are comparable for full, partial and on-the-fly methods (see Fig.(\ref{MPI_time_results})).
Consequently, the MPI communication times combine two intricate effects: the time taken to do a SpMV, and the uneven distribution of Hamiltonian matrix elements among nodes. 
Indeed, processes will not finish their calculations at the same time. Consequently, they start communicating with each other at different times, and they experience additional MPI delays due
to load imbalances effects. Therefore, what we report as "MPI time" must be understood as load imbalance
plus MPI time. Further investigation is then necessary to understand how to mitigate load imbalance, as this will surely decrease MPI communication times.

\section{Conclusion} 
\label{GSM_conclusion}
GSM has been parallelized with the most powerful computing method developed for SM, based on a 2D partitioning of the Hamiltonian matrix. It significantly reduces MPI inter-node communications while allowing to take advantage of  the symmetry of the Hamiltonian. Time-reversal symmetry has also been included in our approach, as such calculations are twice faster for even nuclei. As GSM vectors are scattered among all nodes, memory requirements are very small for the vectors entering the JD method and reorthogonalization of vectors. Moreover, an effective on-the-fly method has been implemented within the 2D partitioning approach, which has been noticed to be slower than the initially developed scheme based on Hamiltonian full storage by a factor of about 5-10, which is still reasonable. The 2D partitioning has been tested with $^8$Be and $^8$He nuclei bearing medium size dimensions, whose calculations resemble those effected in the context of GSM. They have shown the efficiency of our method, despite a load imbalance issue which will be addressed as part of our future work.

Consequently, the implementation of the 2D partitioning method has expanded the matrix dimensions that GSM can treat, and it is now possible to deal with dimensions of the order of tens or hundreds of millions. These calculations will demand for sure the use of very powerful machines, bearing thousands of cores. As the feasibility of these calculations has been demonstrated, it is now possible in a near future to consider very large systems in GSM.

\section{Acknowledgments} 
\label{Acknowledgments}
We thank Dr.~K{\'e}vin Fossez and Prof.~Witek Nazarewicz for useful discussions and comments. N.~Michel wishes to thank Prof.~Furong Xu for a CUSTIPEN visit in Peking University,
as well as Dr.~S.M.~Wang, for letting us use his figure of the Berggren completeness relation.
This work was supported by the US Department of Energy, Office of Science, Office of Nuclear Physics under Awards No.~DE-SC0013365 (Michigan State University), No.~DE-SC0008511 (NUCLEI SciDAC-3 Collaboration), No.~DE-SC0018083 (NUCLEI SciDAC-4 Collaboration); the National Natural Science Foundation of China under Grant No.~11435014;
and No.~DE-SC0009971 (CUSTIPEN: China-U.S. Theory Institute for Physics with Exotic Nuclei), and also supported in part by Michigan State University through computational resources provided by the Institute for Cyber-Enabled Research.
This research used resources of the Oak Ridge Leadership Computing Facility, which is a DOE Office of Science User Facility supported under Contract DE-AC05-00OR22725,
and of the National Energy Research Scientific Computing Center (NERSC), a U.S. Department of Energy Office of Science User Facility operated under Contract No. DE-AC02-05CH11231. \\

This work is licensed under a license Creative Commons Attribution-NonCommercial-NoDerivatives 4.0 International (CC BY-NC-ND 4.0) (https://creativecommons.org/licenses/by-nc-nd/4.0/deed.en\_us).

\bibliography{references}

\end{document}